# Theoretical prediction of MoN$_2$ monolayer as a high capacity electrode material for metal ion batteries

Xiaoming Zhang,[a] Zhiming Yu,[a,b] Shan-Shan Wang,[a] Shan Guan,[a,b] Hui Ying Yang,[*,a] Yugui Yao,[*,b] and Shengyuan A. Yang[*,a]

[a]Research Laboratory for Quantum Materials and Engineering Product Development Pillar, Singapore University of Technology and Design, Singapore 487372, Singapore. E-mail: yanghuiying@sutd.edu.sg; shengyuan_yang@sutd.edu.sg
[b]School of Physics, Beijing Institute of Technology, Beijing 100081, China. E-mail: ygyao@bit.edu.cn

**ABSTRACT:** Benefited from the advantages on environmental benign, easy purification, and high thermal stability, the recently synthesized two-dimensional (2D) material MoN$_2$ shows great potential for clean and renewable energy applications. Here, through first-principles calculations, we show that the monolayered MoN$_2$ is promising to be a high capacity electrode material for metal ion batteries. Firstly, identified by phonon dispersion and exfoliation energy calculations, MoN$_2$ monolayer is proved to be structurally stable and could be exfoliated from its bulk counterpart in experiments. Secondly, all the studied metal atoms (Li, Na and K) can be adsorbed on MoN$_2$ monolayer, with both pristine and doped MoN$_2$ being metallic. Thirdly, the metal atoms possess moderate/low migration barriers on MoN$_2$, which ensures excellent cycling performance as a battery electrode. In addition, the calculated average voltages suggest that MoN$_2$ monolayer is suitable to be a cathode for Li-ion battery and anodes for Na-ion and K-ion batteries. Most importantly, as a cathode for Li-ion battery, MoN$_2$ possesses a comparable average voltage but a 1-2 times larger capacity (432 mA h g$^{-1}$) than usual commercial cathode materials; as an anode for Na-ion battery, the theoretical capacity (864 mA h g$^{-1}$) of MoN$_2$ is 2-5 times larger than typical 2D anode materials such as MoS$_2$ and most MXenes. Finally we also provide an estimation of capacities for other transition-metal dinitrides materials. Our work suggests that the transition-metal dinitrides MoN$_2$ is an appealing 2D electrode materials with high storage capacities.

## 1 Introduction

Featured with extraordinary physical and chemical properties, two-dimensional (2D) materials have been attracting increasing interests in recent years, with a wide range of applications from electronic devices including p-n junctions[1] and field effect transistors[2] to energy storage technologies such as supercapacitors[3] and batteries.[4,5] Especially, the 2D materials based secondary batteries have been a focus of current research because of their high reversible capacity, high power density and excellent cycling stability.[6] Among the secondary batteries, rechargeable Li-ion batteries (LIBs) have received significant success and been widely used in portable electronic devices and electric vehicles since its first commercialization in 1991.[7,8] However, due to their intrinsic safety issues and relatively high cost,[9,10] LIBs face challenges from further developments in large-scale energy storage applications. Therefore, developing new types of batteries is urgently needed. Benefited from the abundant sodium natural deposits and similar storage mechanism with LIBs, Na-ion batteries (NIBs) are expected to be an ideal choice to replace LIBs in future for certain applications.[11,12] Besides LIBs and NIBs, another alkali ion based batteries, namely K-ion batteries (KIBs), have also received considerable attention nowadays.[13,14] It is well known that, the performance of ion batteries strongly depends on their electrode materials. Therefore, developing advanced electrode materials, especially with high energy storage capacity, will continue to be the critical challenge in the battery technologies. So far, a large number of 2D materials, including graphene-related materials,[15,16] metal oxides and metal nitrides,[17,18] transition-metal dichalcogenides (TMDCs),[19,20] MXenes,[21-24] and black phosphorus,[25,26] have been intensively investigated because of their excellent electrochemical performance as battery electrodes.

Quite recently, a new 2D material system, namely the transition-metal dinitrides (TMDNs), has been proposed since the first experimental realization of $MoN_2$.[27] Bulk $MoN_2$ is found to possess a rhombohedral R3*m* structure, which is isotypic with bulk $MoS_2$. As a representative TMDN, high purity bulk $MoN_2$ has been proved stable even at high temperature, and exhibits better activities than traditional $MoS_2$ as catalysts.[27] Especially, a theoretical work[28] by Wu *et al.* predicted that, its monolayer would show intrinsic magnetism with high Cure temperatures and good electronic conductivities, promising for spintronics applications. Beyond these novel properties, how the TMDNs behave as battery electrodes is still unknown, even though, compared with electrodes from TMDCs such as $MoS_2$, TMDNs are more environmentally-friendly, and inherently possess much better electronic conductivity and lower mass density.

In this work, we theoretically study the performance of the newly proposed 2D material system $MoN_2$ as electrodes for metal ion batteries. Through first-principles calculations, we first study the thermal and dynamical stabilities of the $MoN_2$ monolayer, then investigate the metal atoms adsorption and diffusion processes. Our results show that, all the studied alkali atoms Li, Na, K can be adsorbed on $MoN_2$, with the metallic nature of $MoN_2$ maintained. Moreover, for Na and K, we find that the $MoN_2$ monolayer possesses low diffusion barriers and low average open-circuit voltages, which is suitable to be used as anodes for NIBs and KIBs; while for LIBs, the $MoN_2$ monolayer is more likely to serve as a cathode material with a relatively high average open-circuit voltage. Most importantly, we find that the $MoN_2$ monolayer possesses very high alkali ion storage capacities with values of 432 mA h $g^{-1}$ for Li, 864 mA h $g^{-1}$ for Na and 432 mA h $g^{-1}$ for K, which is far superior to typical electrode materials. We also clarify the fundamental origin of the multilayered adsorption behavior of alkali atoms on 2D $MoN_2$. Overall, we propose for the first time that the transition-metal dinitrides, especially $MoN_2$ can be a promising high capacity electrode material for the next-generation metal ion batteries.

## 2  Computational methods

The first-principles calculations are carried out in the framework of density functional theory (DFT), as implemented in the Vienna ab initio simulation package (VASP).[29] The generalized gradient approximation (GGA) as formulated by Perdew−Burke−Ernzerhof (PBE) functional is employed for the exchange-correlation potential.[30,31] The cutoff energy is chosen as 500 eV for the plane-wave expansion of valence electron wave functions. To avoid artificial interaction between two isolated monolayers, a vacuum spacing of 20 Å is built. For geometrical optimization, a $21 \times 21 \times 1$ Monkhorst-Pack k-point mesh[32] is used for the unit cell of the $MoN_2$ monolayer. Meanwhile, DFT-D2 method[33] is employed to describe the long-range van der Waals interactions. Both the atomic positions and lattice constants are optimized, and the convergence criteria for energy and force are set to be $10^{-5}$ eV and 0.01 eV Å$^{-1}$, respectively.

To investigate the dynamical stability of the $MoN_2$ monolayer, the phonon spectra are calculated by using the PHONOPY package.[34,35] The amount of charge transfer between alkali atoms and $MoN_2$ monolayer is estimated by using the Bader charge method.[36-38] The climbing-image nudged elastic band (CI-NEB) method[39] is applied to determine the optimized diffusion path of alkali atoms on the surface of the $MoN_2$ monolayer. In this work, the CI-NEB calculations are performed using 11 images, including the initial and final structural configurations. Then the geometry and energy of the images are fully relaxed until the largest norm of the force orthogonal to the path become smaller than 0.01 eV Å$^{-1}$.

## 3  Results and discussions

### 3.1  Structure and stability of the $MoN_2$ monolayer

We first determine the most energetically stable structure for $MoN_2$ monolayer. Under similar consideration of $MoS_2$, we assign three types of structures for $MoN_2$ (namely 1H,

1T and 1T'). The comparison of total energy shows that the 1H one is the ground-state structure for MoN$_2$ monolayer. As shown in Fig. 1(a), on the top view, the 1H-MoN$_2$ exhibits a hexagonal structure with Mo and N atoms occupying the two sites of a honeycomb lattice. On the side view, as shown in Fig. 1(b), each bare MoN$_2$ monolayer contains a triple layer of atoms stacked in the sequence of N−Mo−N. During the structural optimization of the MoN$_2$ monolayer, we take into account both ferromagnetic (FM) and nonspin polarized (NSP) states. The results show that FM is the ground state with 0.21 eV/unit cell lower in energy than the NSP state. The optimized lattice constants are determined to be a=b=2.99 Å with a N-Mo interatomic distance of 2.05 Å. Then we turn to the electronic structures of the MoN$_2$ monolayer at the ground state. The calculated total and projected density of states (DOS) are shown in Fig. 2(b). In the total DOS, the split of the electronic structures between the spin up and the spin down channels manifests the FM character of the MoN$_2$ monolayer. Drawn from the projected DOS, the spin polarization in the MoN$_2$ monolayer mostly originates from the *p* orbitals of N atom and *d* orbitals of Mo atom. The calculated magnetic moments of N and Mo atoms are 0.42 μ$_B$ and 0.14 μ$_B$, respectively. Our results are consistent with former calculations.[28] More importantly, MoN$_2$ monolayer is metallic with sizable DOS at Fermi level, making it possible to be used as a battery electrode material.

Even though MoN$_2$ has been successfully synthesized in bulk condition,[27] it is still necessary to investigate the stability of monolayered MoN$_2$. Firstly, we calculate formation energy of MoN$_2$ monolayer by using the following equation,

$$E_F = E_{MoN_2} - E_{Mo} - E_{N_2} \qquad (1)$$

where $E_F$ is the formation energy, and $E_{MoN_2}$, $E_{Mo}$ and $E_{N_2}$ are the energies of MoN$_2$ monolayer, pure Mo metal and nitrogen, respectively. The calculated formation energy is -1.81 eV/unit cell, indicating that the MoN$_2$ monolayer is thermodynamically stable. Moreover, we investigate the dynamical stability of MoN$_2$ monolayer through phonon dispersion calculations. As shown in Fig. 2(a), the phonon spectra exhibit no imaginary frequency with the optical and acoustical branches well separated. Meanwhile, in the vicinity of Γ point, the two in-plane acoustic phonons exhibit linear dispersion while the out-plane (ZA) acoustic branch displays a quadratic dispersion, manifesting a typical phonon dispersion character of layered materials.[40] In this respect, the given structure of MoN$_2$ monolayer is dynamically stable. Furthermore, we investigate the possibility of exfoliating the MoN$_2$ monolayer from its bulk by evaluating the exfoliation energy:

$$E_E = E_{Bulk}/N - E_{Mono} \qquad (2)$$

where $E_E$ is the exfoliation energy, $E_{Bulk}$ and $E_{Mono}$ are the total energies of MoN$_2$ in bulk and monolayer, and $N$ is the number of layers. The exfoliation energy of MoN$_2$ is calculated to be 0.17 eV/unit cell, which is very close to MoS$_2$ (0.16 eV/unit cell).[28] It suggests that by following similar experimental technologies (such as mechanical exfoliation method) as applied to MoS$_2$,[41,42] MoN$_2$ monolayer or few-layers may also be obtained in experiments. From the calculations on formation energy, phonon spectra and exfoliation energy, we conclude that the MoN$_2$ monolayer is structural stable and is promising to be realized in experiments.

### 3.2 Alkali atom adsorption on the MoN$_2$ monolayer

To study the nature of alkali (Li, Na, and K) atom adsorption on the MoN$_2$ monolayer, we first determine the most favorable adsorption site for a single alkali atom. Herein, four possible adsorption sites (denoted as S1-S4) are considered, as depicted in Fig. 1(c). To obtain more accurate results, we apply full structural optimizations for all the adsorption sites under both FM and NSP states. A 2 ×2 supercell of MoN$_2$ monolayer is adopted in the calculation. The most favorable adsorption configurations can be identified by comparing their adsorption energies, which are defined as,

$$E_{Ad} = E_{M+MoN_2} - E_{MoN_2} - E_A \qquad (3)$$

where $E_{Ad}$ is the adsorption energy, $E_{MoN_2}$ and $E_{M+MoN_2}$ are the total energies of MoN$_2$

supercell before and after adsorptions, and $E_A$ represents the energy per atom in bulk alkali (Li, Na, and K) metals.

As shown in Fig. 3(a), the calculated adsorption energies for Li, Na, and K manifest similar features: firstly, for all the adsorption sites and both FM and NSP states, $MoN_2$ exhibits negative Li/Na/K adsorption energies, indicating that the alkali atoms prefer to adsorb on the $MoN_2$ monolayer rather than forming metal clusters; secondly, the adsorption energies under the NSP state are larger than the FM state, which shows that the $MoN_2$ monolayer favors the NSP state after adsorptions of alkali atoms; thirdly, among the four adsorption sites (S1-S4), S3 site is found to have the lowest adsorption energy. Therefore, for all the alkali Li, Na, and K atoms, the adsorption above the center of the holes of honeycomb lattice (S3 site) under the NSP state serves as the most favorable configuration. To be noted, different from the former nonmagnetic systems such as TMDCs and MXenes,[19-24] in magnetic $MoN_2$ monolayer, one should carefully examine not only the various adsorption sites but also the different magnetic states to correctly determine the most favorable absorption configuration.

To get further insights into the adsorption process, we investigate the charge transfers by performing Bader charge analysis. At the ground state (NSP), we find that the alkali atoms almost donate all their outmost $s$ electrons to the $MoN_2$ monolayer, and the transferred charges for Li, Na and K are calculated to be 0.89 e, 0.86 e, and 0.84 e, respectively. These charge transfers suggest that the alkali ions are in the cationic state and chemically adsorbed onto the $MoN_2$ monolayer. For comparison, we also calculate the charge transfers under FM state. We find that the transferred charges for Li, Na and K almost show no difference between FM and NSP state. Therefore, from the charge transfer point of view, the magnetism in $MoN_2$ monolayer has little influence on the adsorption process of alkali atoms.

It is well known that, good conductivity is very critical for an electrode material. On considering this, we study the electronic structures of the $MoN_2$ monolayer after the alkali atom adsorption. Fig. 3(b) shows the total DOS at the optimized configuration for Li, Na, and K adsorption. Obviously, the DOS profiles near the Fermi level are very similar upon Li, Na, and K adsorptions, with the alkali atoms provide electrons to the $MoN_2$ monolayer. More importantly, the calculated DOS clearly exhibits a metallic character, which ensures good electronic conduction when used as battery electrodes.

### 3.3 Alkali atom diffusion on the $MoN_2$ monolayer

The rate performance of a battery electrode is mainly determined by the mobility of the intercalating ions, so in the following we estimate the diffusion behavior of alkali atoms on the surface of the $MoN_2$ monolayer by using the CI-NEB method. As shown in Fig. 4(a), we examine three possible migration pathways along the high symmetry line between two neighboring favorable adsorption sites, where three different initial pathways are denoted as P1, P2 and P3. For Li and Na atoms, the pathway of P2, which passes through the high symmetry site above the Mo atom, is found to possess the lowest diffusion barrier; while for K atom, the optimized pathway turns out to be P3. The corresponding diffusion energy profiles along the optimized pathway for Li, Na and K are shown in Fig. 4(b), (c) and (d), respectively.

Among the three alkali atoms, Li possesses the highest diffusion barriers with a value of 0.78 eV. Although higher than most 2D materials such as graphene (0.33 eV) [43] and $MoS_2$ (0.25 eV),[19] the value is comparable to many well-studied electrodes materials including silicon (0.57 eV),[44,45] and $TiO_2$-based polymorphs (~0.65 eV).[46,47] Recently, Guo et al.[48] found that the Li diffusion barrier in phosphorene can be significantly reduced after introducing specific vacancy defects. Similar strategy may also apply for $MoN_2$ monolayers on the purpose of improving the Li ion mobility. In comparison, the diffusion barriers for Na (0.56 eV) and K (0.49 eV) are both lower than 0.6 eV, which ensures moderate/high Na and K diffusion mobility on the surface of $MoN_2$ monolayer. From the calculated diffusion barriers, excellent charge-discharge rate can be expected for $MoN_2$ monolayers when used as electrodes for NIBs and KIBs.

## 3.4 Theoretical storage capacity and open circuit voltage of the MoN$_2$ monolayer

Here we evaluate the open circuit voltage (OCV) and theoretical Li/Na/K storage capacity of MoN$_2$ monolayer, which are the most crucial parameters to determine the performance of an electrode material. To explore the maximum storage capacity for Li, Na and K, we consider the adsorption on both sides of MoN$_2$ monolayer. Here a 2 × 2 supercell of MoN$_2$ monolayer is adopted in the calculation. The charge-discharge process of MoN$_2$ monolayer can be described as the common half-cell reaction vs. A/A$^+$:

$$MoN_2 + xA^+ + xe^- \longleftrightarrow MoN_2A \quad (4)$$

where A represents the alkali element of Li, Na or K. For the given reaction, we can estimate the average OCV ($V_{ave}$) based on the energy difference expressed by following equation:

$$V_{ave} = (E_{MoN_2} + xE_A - E_{MoN_2A_x})/xye \quad (5)$$

where $E_{MoN_2}$ and $E_{MoN_2A_x}$ are the total energies of MoN$_2$ monolayer before and after A$^+$ intercalation, $E_A$ is the energy per atom in bulk alkali (Li, Na, and K), and y represents the electronic charge of alkali ions in the electrolyte (here, y =1).

Before computing $V_{ave}$, we first evaluate the maximum storage of Li, Na and K on the MoN$_2$ monolayer. As mentioned above, the most stable adsorption site for Li, Na and K all locates above the center of the N-Mo honeycomb lattice (S3 site), therefore we apply the first Li/Na/K intercalation layer at S3 site. For the second layer, we examine all the possibilities by successively assigning the adatoms at S1, S2 and S4 sites to determine the most stable one. We can investigate the Li/Na/K storage performance of MoN$_2$ monolayer by calculating the average adsorption energy ($E_{ave}$) layer by layer, which is defined as:

$$E_{ave} = \left(E_{MoN_2A_{8n}} - E_{MoN_2A_{8(n-1)}} - 8E_A\right)/8 \quad (6)$$

where $E_{MoN_2A_{8n}}$ and $E_{MoN_2A_{8(n-1)}}$ are the total energies of MoN$_2$ with *n* and *n-1* adsorption layers respectively, $E_A$ represents the energy per atom in bulk alkali Li, Na, and K, and the number of "8" corresponds to eight adatoms for each layer.

For the first adsorption layer, the calculated average adsorption energy is -2.78 eV/atom for Li, -1.64 eV/atom for Na, and -1.09 eV/atom for K. These large negative adsorption energies ensure good intercalation stabilities of Li/Na/K ions on MoN$_2$ monolayer. For the second adsorption layer, we find the configuration with adatoms at S2 site (see Fig. 1c) possesses the lowest average adsorption energies, which are calculated to be 0.18 eV/atom for Li, -0.02 eV/atom for Na and 1.40 eV/atom for K. Obviously, only the second layer of Na is energetically stable, while those of Li and K are unstable. To be noted here, the second-layer for Na possesses a low absorption energy (-0.02 eV/atom), indicating a weak multilayered adsorption. However, the value is comparable or even larger than typical electrode materials, such as the adsorption energies of Na on Ca$_2$N (-0.003 eV/atom)[17] and GeS (-0.02 eV/atom)[49]; and Li on Nb$_2$C (-0.02 eV/atom)[50] and Mo$_2$C (-0.01 eV/atom).[21] Therefore, it is practice to use it for the estimation of theoretical maximum capacity. Even though the adsorption interactions in experiments can be affected by many complicated factors, such as the quality and morphology of electrodes, the type of electrolyte and its concentration, and so on, the theoretical DFT result is still widely accepted as useful for guiding and interpreting experimental studies. Considering the of absorption energy, the MoN$_2$ monolayer can accommodate up to 1 layer of Li, 2 layers of Na and 1 layer of K, respectively. Then we clarify the physical origin of the diverse adsorption behaviors of alkali atoms on the MoN$_2$ monolayer. In equation (6), the sign of the layered $E_{ave}$ determines the metal atoms in corresponding layer prefer to chemically bound to the host material (negative $E_{ave}$) or to form metal clusters themselves (positive $E_{ave}$). In Fig. 5(a), (b) and (c), we display the electron localization functions (ELF) of the (110) section of the MoN$_2$ monolayer with two layers of (a) Li, (b)

Na and (c) K, respectively. For the second layer of Li, in the ELF graph (see Fig.5a) we find sufficient concentration of electrons between Li atoms in the outer layer, indicating strong bounding among them. As a result, the second layer of Li atoms are more likely to form Li clusters rather than adsorbing on the $MoN_2$ monolayer. Unlike Li, in the ELF graph of K (see Fig.5b), the electrons concentration no longer locates between atoms in the layer, leading to a weaker bounding among these K atoms. Even though, due to the extremely large distance (6.23 Å) between the second K layer and the host $MoN_2$, the corresponding interactions are too weak to trigger their chemical bounding. However, for the second layer of Na ( see Fig. 5b), it possesses a weaker bounding strength among Na atoms in the outer layer than that of Li, and also a much smaller distance of Na-$MoN_2$ than that of K-$MoN_2$ (4.70 Å *vs.* 6.23 Å), which gives rise to the possibility of the multilayered adsorption behavior of Na on the $MoN_2$ monolayer. Such a qualitative competition mechanism is found to well agrees with the calculated adsorption energies.

Here we continue to evaluate the maximum Li/Na/K storage capacity of the $MoN_2$ monolayer. As mentioned above, the $MoN_2$ monolayer can accommodate up to 1 layer of Li, 2 layers of Na and 1 layer of K, corresponding to chemical stoichiometry of $MoN_2Li_2$, $MoN_2Na_4$ and $MoN_2K_2$, respectively. The maximum capacity ($C_M$) can be obtained by the following equation,

$$C_M = xF / M_{MoN_2} \quad (7)$$

where $x$ represents the number of electrons involving the electrochemical process, $F$ derives is the Faraday constant with the value of 26.8 A h mol$^{-1}$, and $M_{MoN_2}$ is the mass of $MoN_2$. The maximum capacity of Li, Na and K are calculated to be 432, 864, and 432 mA h g$^{-1}$, respectively. Meanwhile, from equation (5), we can obtained the average OCV, which is 3.64 V for Li, 1.91 V (one layer) to 0.96 V (double layers) for Na, and 1.11 V for K. Compared with Li, we find Na possesses a dramatically lower average OCV, which has also been claimed in other electrode materials.[21,22,51] Recently, evidenced by sufficient calculations, Liu *et al.* have ascribed such phenomenon to the relatively weaker Na chemical binding to the substrates, originating from the competition between the ionization of the metal atom and the ion–substrate coupling.[52] Considering the high average OCV of Li, the $MoN_2$ monolayer can be used as a cathode material for LIB with a theoretical capacity of 432 mA h g$^{-1}$. Comparisons of maximum capacity and average OCV between the $MoN_2$ monolayer and some widely investigated cathode materials are shown in the Supplementary Information (SI). To our satisfied, we find the $MoN_2$ monolayer shows comparable average OCV but 1-2 times larger in capacity than the commercial cathode materials including $LiCoO_2$,[53,54] $LiMn_2O_4$ and so on.[55-58] As for the S cathode, it possesses an extremely high capacity (1675 mA h g$^{-1}$),[59,60] but suffers from several disadvantages for cathode applications including low average Li/Li$^+$ potential (2.15 V), volume explosion, serious capacity fading, and also bad electrical conductivity.[61,62] Overall, the $MoN_2$ monolayer can be a good choice for a LIBs cathode material with combining characteristics, such as excellent electrical conductivity, high charging voltage and high energy storage capacity. For NIBs and KIBs, the $MoN_2$ monolayer is more likely to serve as an anode material because of the relatively low average OCV (0.96 V *vs.* Na/Na$^+$ and 1.11 V *vs.* K/K$^+$). Especially, as an anode for NIBs, the theoretical capacity in $MoN_2$ (864 mA h g$^{-1}$) can be 2-5 times larger than most 2D anodes materials, such as TMDCs including $MoS_2$ (146 mA h g$^{-1}$),[63] and $NbS_2$ (263 mA h g$^{-1}$),[64] and MXenes including $Nb_2C$ (271 mA h g$^{-1}$),[50] $Mo_2C$ (132 mA h g$^{-1}$),[21,22] and soon,[65,66] and also much larger than the recently reported GeS nanosheet (512 mA h g$^{-1}$).[49] The obtained high Na-ion capacity in the $MoN_2$ monolayer mainly originates from its low atomic weight and multilayered adsorption ability. Besides higher in capacity, $MoN_2$ as battery electrodes also possesses advantages in: (1) compared with TMDCs, $MoN_2$ is more cost effective, environmentally friendly and also possesses better electrical conductivity; (2) compared with MXenes, $MoN_2$ is easier in fabrication and more stable, while the surfaces of MXenes are not stable and often easily passivated by F/OH functional groups during preparations,[67,68] which will greatly deteriorate the stability and their electrochemical performances as battery electrodes.

In TMDCs, the transition-metal element is not limited to 4$d$ (Y, Zr, Nr, Mo etc.) transition-metals, but may also from 3$d$ (Ti, V, Cr, etc.) and 5$d$ (Hf, Ta, W, etc.) categories. Under similar considerations, TMDNs (here denoted as MN$_2$) may also form a big family of 2D materials like TMDCs. Follow equation (7), here we make a rough approximate estimation of the theoretical capacities of TMDNs as battery electrodes. For single-layered adsorption, the capacities are evaluated to be ~260 mA h g$^{-1}$ for 5$d$-MN$_2$, ~430 mA h g$^{-1}$ for 4$d$-MN$_2$, and ~680 mA h g$^{-1}$ for 3$d$-MN$_2$. By considering the possibility of multilayered adsorption, the capacities obtained above could even double, especially the capacities in 3$d$-MN$_2$ can reach as high as ~1200 mA h g$^{-1}$. Our results highlight the promise of TMDNs as high capacity electrodes materials for the next-generation batteries.

## 4 Summary

In this work, by using the MoN$_2$ monolayer as an example, we investigate the potential of TMDNs as electrode materials for metal ion batteries on the basis of first-principles calculations. Through the calculations on formation energy, phonon spectra and exfoliation energy, we confirmed that the MoN$_2$ monolayer is structurally stable and promising to be exfoliated from its bulk counterpart in experiments. Our results show all the studied alkali atoms (Li, Na, K) possess negative adsorption energies , indicating their stable adsorptions on the MoN$_2$ monolayer. Moreover, the metallic character of the MoN$_2$ monolayer both before and after adsorptions ensures excellent electronic conductivity when used as battery electrodes. The most energetically favorable diffusion pathways of alkali atoms are indentified. The results indicate that the MoN$_2$ monolayer can possess good charge-discharge rates as battery electrodes. On considering the calculated average OCV and ion-storage capacity, our work reveals that: (1) the MoN$_2$ monolayer is more likely to be used as cathodes for LIBs and anodes for NIBs and KIBs; (2) As a cathode material for LIBs, the MoN$_2$ monolayer manifests a comparable average OCV (3.64 V *vs.* Li/Li$^+$) but a much larger capacity (432 mA h g$^{-1}$) than usual commercial cathode materials; (3) the MoN$_2$ monolayer shows a multilayered adsorption ability for Na, and the theoretical Na-ion capacity (864 mA h g$^{-1}$) is 2-5 times larger than typical 2D anode materials such as MoS$_2$ and most MXenes. Furthermore, our results suggest that, besides MoN$_2$, many other TMDNs are also promising to be used as battery electrodes with high storage capacities.

## Acknowledgments

This work is supported by Singapore Ministry of Education Academic Research Fund Tier 2 (MOE2015-T2-1-150), SUTD-SRG-EPD-2013062. Yugui Yao is supported by the MOST Project of China (No. 2014CB920903), the National Natural Science Foundation of China (Grant Nos. 11574029, and 11225418).

## References

1 M.-Y. Li, Y. Shi, C.-C. Cheng, L.-S. Lu, Y.-C. Lin, H.-L. Tang, M.-L. Tsai, C.-W. Chu, K.-H. Wei, J.-H. He, W.-H. Chang, K. Suenaga and L.-J. Li, *Science*, 2015, **349**, 524–528.

2 B. Radisavljevic, A. Radenovic, J. Brivio, V. Giacometti and A. Kis, *Nat. Nanotechnol.*, 2011, **6**, 147-150.

3 M. Acerce, D. Voiry and M. Chhowalla, *Nat. Nanotechnol.*, 2015, **10**, 313–318.

4 B. Dunn, H. Kamath and J.-M. Tarascon, *Science*, 2011, **334**, 928−935.

5 G. Zhou, F. Li and H.-M. Cheng, *Energy Environ. Sci.*, 2014, **7**,1307–1338.

6 H. Chen, T. N. Cong, W. Yang, C. Tan, Y. Li and Y. Ding, *Nat. Sci.*, 2009, **19**, 291−312.

7 J. -M. Tarascon and M. Armand, *Nature*, 2001, **414**, 359−367.

8 D. P. Dubal, O. Ayyad, V. Ruiz and P. Gomez-Romero, *Chem. Soc. Rev.*, 2015, **44**, 1777

−1790.

9 J.-M. Tarascon, *Nat. Chem.*, 2010, **2**, 510–510.

10 J. B. Goodenough and K.-S. Park, *J. Am. Chem. Soc.*, 2013, **135**, 1167–1176.

11 H. M. Liu, H. S. Zhou, L. P. Chen, Z. F. Tang and W. S. Yang, *J. Power Sources*, 2011, **196**, 814−819.

12 Y. H. Lu, L. Wang, J. G. Cheng and J. B. Goodenough, *Chem. Commun.*, 2012, **48**, 6544−6546.

13 A. Eftekhari, *J. Power Sources*, 2004, **126**, 221–228.

14 C. D Wessells, R. A Huggins and Y. Cui, *Nat. Commun.*, 2011, **2**, 550.

15 K. S. Novoselov, A. K. Geim, S. V. Morozov, D. Jiang, M. I. Katsnelson, I. V. Grigorieva, S. V. Dubonos and A. A. Firsov, *Nature*, 2005, **438**, 197–200.

16 E. Yoo, J. Kim, E. Hosono, H. shen Zhou, T. Kudo and I. Honma, *Nano Lett.*, 2008, **8**, 2277–2282.

17 J. Hu, B. Xu, S. A. Yang, S. Guan, C. Ouyang and Y. Yao, *ACS Appl. Mater. Interfaces*, 2015, **7**, 24016−24022.

18 S. Deng, L. Wang, T. Hou and Y. Li, *J. Phys. Chem. C*, 2015, **119**, 28783–28788.

19 L. David, R. Bhandavat and G. Singh, *ACS Nano*, 2014, **8**, 1759−1770.

20 R. Bhandavat, L. David and G. Singh, *J. Phys. Chem. Lett.*, 2012, **3**, 1523–1530.

21 Q. Sun, Y Dai, Y. Ma, T. Jing, W. Wei and B. Huang, *J. Phys. Chem. Lett.*, 2016, **7**, 937−943.

22 D. Çakır, C.Sevik, O. Gülserenc and F. M. Peetersa, *J. Mater. Chem. A*, 2016, **4**, 6029-6034.

23 D. Er, J. Li, M. Naguib, Y. Gogotsi and V. Shenoy, *ACS Appl. Mater. Interfaces*, 2014, **6**, 11173−11179.

24 C. Eames and M. S. Islam, *J. Am. Chem. Soc.*, 2014, **136**, 16270–16276.

25 J. Sun, H.-W. Lee, M. Pasta, H. Yuan, G. Zheng, Y. Sun, Y. Li and Y. Cui, *Nat. Nanotechnol.*, 2015, **10**, 980–985.

26 W. Li, Y. Yang, G. Zhang and Y.-W. Zhang, *Nano Lett.*, 2015, **15**, 1691−1697.

27 S. M. Wang, H. Ge, S. L. Sun, J. Z. Zhang, F. M. Liu, X. D. Wen, X. H. Yu, L. P. Wang, Y. Zhang, H. W. Xu, J. C. Neuefeind, Z. F. Qin, C. F. Chen, C. Q. Jin, Y. W. Li, D. W. He and Y. S. Zhao, *J. Am. Chem. Soc.*, 2015, **137**, 4815.

28 F. Wu, C. Huang, H. Wu, C. Lee, K. Deng, E. Kan and P. Jena, *Nano Lett.*, 2015, **15**, 8277−8281.

29 G. Kresse and D. Joubert, *Phys. Rev. B: Condens. Matter Mater. Phys.*, 1999, **59**, 1758−1775.

30 J. P. Perdew, K. Burke, Y. and Wang, *Phys. Rev. B: Condens. Matter Mater. Phys.*, 1996, **54**, 16533−16539.

31 J. P. Perdew, K. Burke and M. Ernzerhof, *Phys. Rev. Lett.*, 1996, **77**, 3865−3868.

32 H. J. Monkhorst and J. D. Pack, *Phys. Rev. B*, 1976, **13**, 5188−5192.

33 S. J. Grimme, *Comput. Chem.*, 2006, **27**, 1787-1799.

34 A. Togo, F. Oba and I. Tanaka, *Phys. Rev. B: Condens. Matter Mater. Phys.* 2008, **78**, 134106.

35 X. Gonze and C. Y. Lee, *Phys. Rev. B*, 1997, **55**, 10355−10368.

36 W. Tang, E. Sanville and G. Henkelman, *J. Phys.: Condens. Matter*, 2009, **21**, 084204.

37 E. Sanville, S. D. Kenny, R. Smith and G. Henkelman, *J. Comput. Chem.*, 2007, **28**, 899.

38 G. Henkelman, A. Arnaldsson and H. Jonsson, *Comput. Mater. Sci.*, 2006, **36**, 254.

39 G. Henkelman, B. P. Uberuaga and H. Jonsson, *J. Chem. Phys.*, 2000, **113**, 9901–9904.

40 H. Zabel, *J. Phys.: Condens. Matter*, 2001, **13**, 7679.

41 B. Radisavljevic, A. Radenovic, J. Brivio, V. Giacometti and A. Kis, *Nat. Nanotechnol.*, 2011, **6**, 147-150.

42 Z. Yu, Y. Pan, Y. Shen, Z. Wang, Z. Ong, T. Xu, R. Xin, L. Pan, B. Wang, L. Sun, J. Wang, G. Zhang, Y. Zhang, Y. Shi and X. Wang, *Nat. Commun.*, 2014, **5**, 5290.

43 E. Pollak, B. Geng, K.-J. Jeon, I. T. Lucas, T. J. Richardson, F. Wang and R. Kostecki,

*Nano Lett.*, 2010, **10**, 3386−3388.

44 O. I. Malyi, T. L. Tan and S. Manzhos, *Appl. Phys. Express*, 2013, **6**, 027301.

45 V. V. Kulish, O. I. Malyi, M. F. Ng, P. Wu and Z. Chen, *RSC Adv.*, 2013, **3**, 4231–4236.

46 M. V. Koudriachova, N. M. Harrison and S. W. Leeuw, *Phys. Rev. Lett.*, 2001, **86**, 1275-1278.

47 M. Wagemaker, R. van de Krol, A. P. Kentgens, A. A. Van Well and F. M. Mulder, *J. Am. Chem. Soc.*, 2001, **123**, 11454–11461.

48 G. C. Guo, X. L. Wei, D. Wang, Y. Luo and L. M. Liu, *J. Mater. Chem. A*, 2015, **3**, 11246–11252.

49 F. Li, Y. Qu and M. Zhao, *J. Mater. Chem. A*, 2016, **4**, 8905-8912.

50 J. Hu, B. Xu, C. Ouyang, Y. Zhang and S. A. Yang, *RSC Adv.*, 2016, **6**, 27467–27474.

51 S. P. Ong, V. L. Chevrier, G. H., A. Jain, C. Moore, S. Kim, X. Ma and G. Ceder, *Energy Environ. Sci.*, 2011, **4**, 3680-3688.

52 Y. Liu, B. V. Merinov and W. A. Goddard III, *PNAS*, 2016, **113**, 3735–3739.

53 K. Mizushima, P. C. Jones, P.J. Wiseman and J. B. Goodenough, *Mater. Res. Bull.*, 1980, **15**, 783–789.

54 T. Ohzuku, A. Ueda, M. Nagayama, Y. Iwakashi and H. Komori, *Acta*, 1993, **38**, 1159–1167.

55 T. Ohzuku and Y. Makimura, *Chem. Lett.*, 2001, **30**, 642–643.

56 J. M. Tarascon, W. R. McKinnon, F. Coowar, T. N. Bowmer, G. Amatucci and D.Guyomard, *J. Electrochem. Soc.*, 1994, **141**, 1421–1431.

57 A. K. Padhi, K. S. Nanjundaswamy, and J. B. Goodenough, *J. Electrochem. Soc.*, 1997, 144, 1188-1197 .

58 A. Yamada, S. C. Chung, and K. Hinokuma, *J. Electrochem. Soc.*, 2001, **148**, A224-A229.

59 E. Peled, A. Gorenshtein, M. Segal and Y. Sternberg, *J. Power Sources*, 1989, **26**,

269-271.

60 P. G. Bruce, S. A. Freunberger, L. J. Hardwick and J.-M. Tarascon, *Nat. Mater.*, 2012, **11**, 19-29.

61 J. Hu, B. Xu, C. Ouyang, S. A. Yang and Y. Yao, *J. Phys. Chem. C*, 2014, **118**, 24274−24281.

62 X. Ji and L.F. Nazar, *J. Mater. Chem.*, 2010, **20**, 9821-9826.

63 S. E. Cheon, S. S. Choi, J. S. Han, Y. S. Choi, B. H. Jung and H. S. Lim, *J. Electrochem. Soc.*, 2004, **151**, 2067-2073.

64 M. Mortazavi, C. Wang, J. K. Deng, V. B. Shenoy and N. V. Medhekar, *J. Power Sources*, 2014, **268**, 279−286.

65 E. Yang, H. Ji and Y. Jung, *J. Phys. Chem. C*, 2015, **119**, 26374−26380.

66 Y. Xie, Y. DallAgnese, M. Naguib, Y. Gogotsi, M. W. Barsoum, H. L. Zhuang and P. R. Kent, *ACS Nano*, 2014, **8**, 9606−9615.

67 M. Naguib, M. Kurtoglu, V. Presser, J. Lu, J. J. Niu, M. Heon, L. Hultman, Y. Gogotsi and M. W. Barsoum, *Adv. Mater.*, 2011, **23**, 4248−4253.

68 O. Mashtalir, M. Naguib, V. N. Mochalin, Y. Dall'Agnese, M. Heon, M. W. Barsoum and Y. Gogotsi, *Nat. Commun.*, 2013, 4, 1−7.

**Figures and captions:**

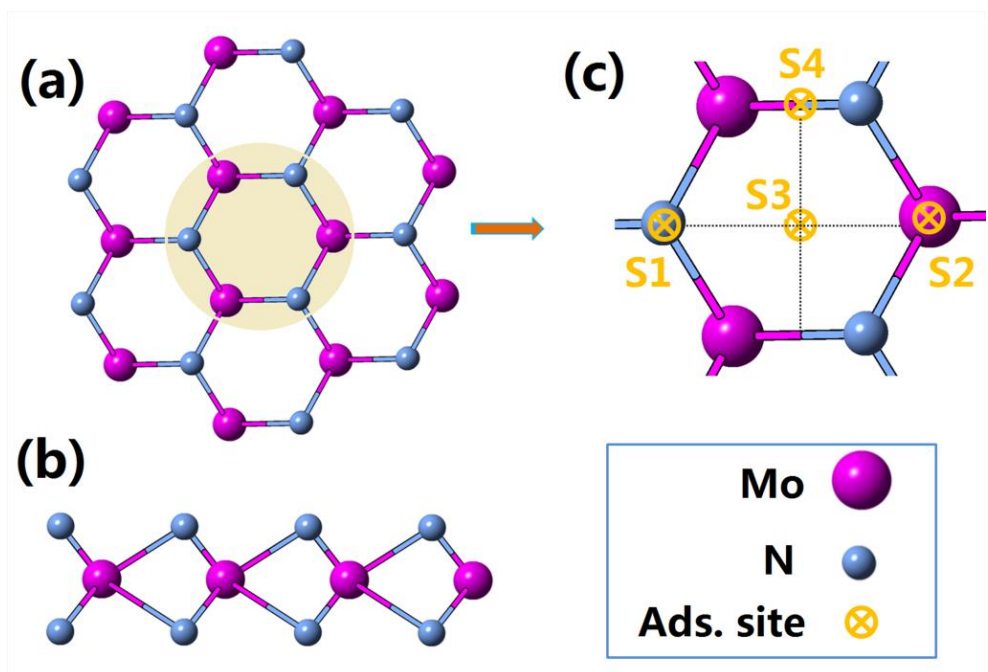

**Fig. 1** Atomic structures of the MoN$_2$ monolayer in (a) top and (b) side view. (c) The considered adsorption sites on the surface of a MoN$_2$ monolayer (top view).

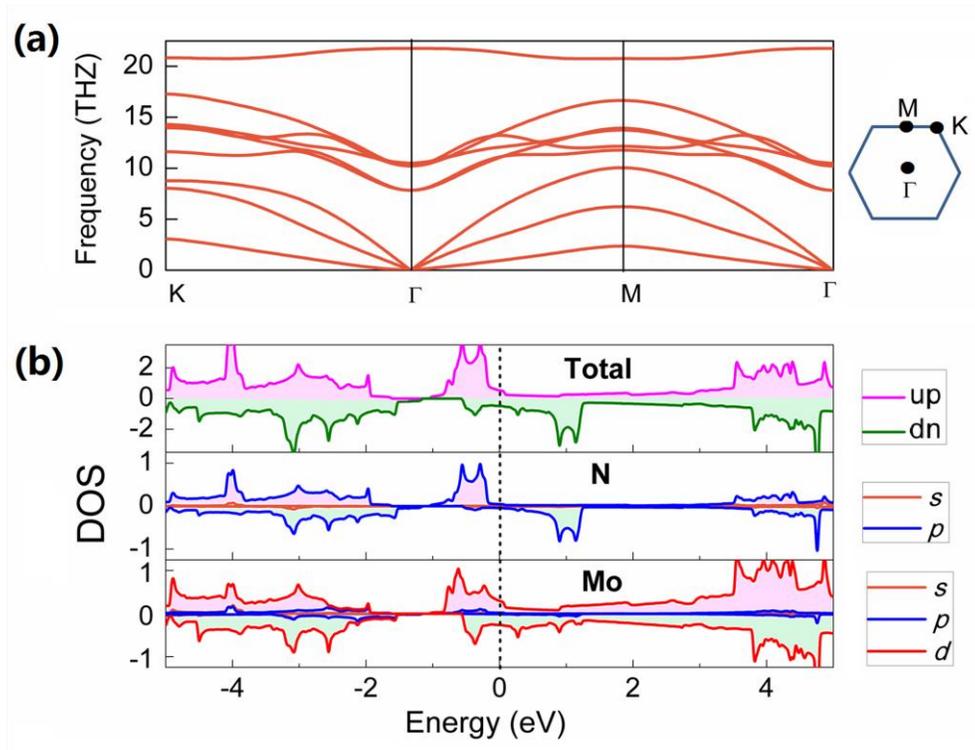

**Fig. 2** (a) The phonon dispersion curves of the fully relaxed MoN$_2$ monolayer. (b) The total and projected density of states of the MoN$_2$ monolayer.

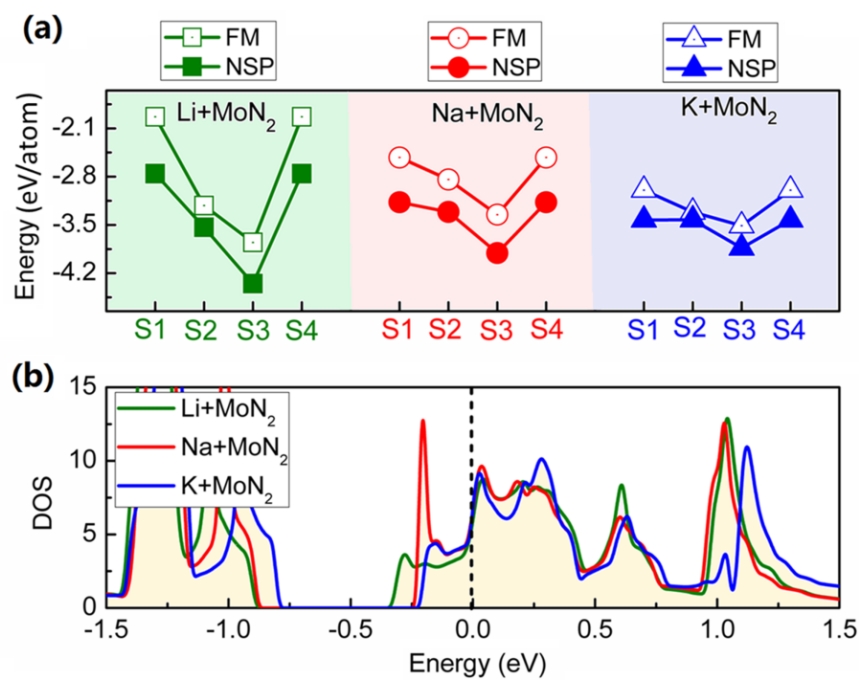

**Fig. 3** (a) Comparison of the Li/Na/K adsorption energies with ferromagnetic (FM) and non-spin polarized (NSP) configurations, where four possible adsorption sites on the surface of the $MoN_2$ monolayer are considered. (b) The total density of states (DOS) of the $MoN_2$ monolayer at the optimized adsorption configuration of Li, Na and K.

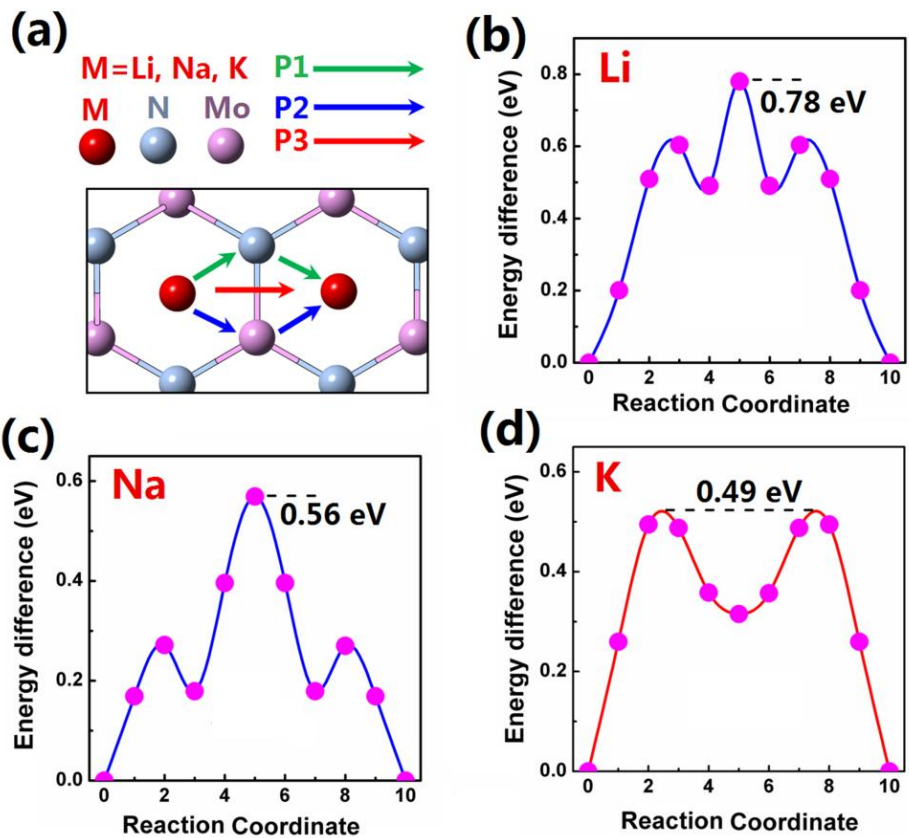

**Fig. 4** (a) Schematic representations of the three possible migration paths of Li/Na/K diffusion on the MoN$_2$ monolayer (top view). (b), (c) and (d) are the diffusion barrier profiles under the optimized pathway for Li, Na and K, respectively.

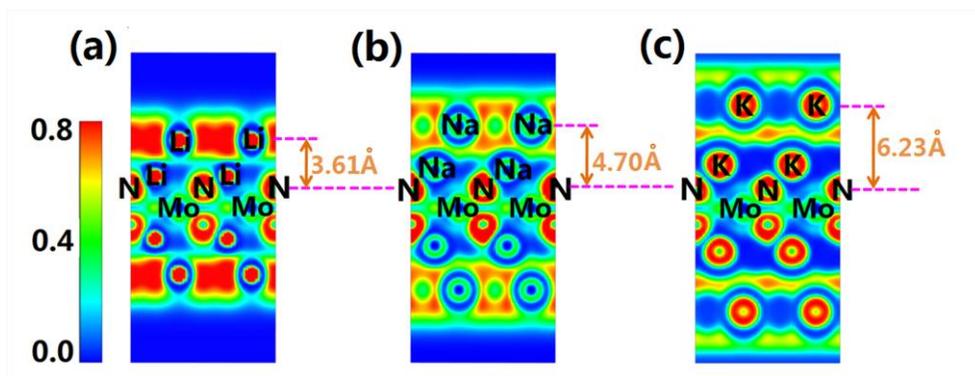

**Fig. 5** The electron localization functions of the (110) section of the $MoN_2$ monolayer with two layers of (a) Li, (b) Na and (c) K.